\documentclass[12pt,preprint]{aastex}
\slugcomment{
}

\shorttitle{Prior Emission Model of X-ray Plateau Phase}
\shortauthors{Ryo Yamazaki}

\begin{document}

\title{
Prior Emission Model for X-ray Plateau Phase 
of Gamma-Ray Burst Afterglows
}

\author{Ryo Yamazaki\altaffilmark{1}}
\affil{Department of Physical Science, Hiroshima University,
Higashi-Hiroshima 739-8526, Japan}
\email{ryo@theo.phys.sci.hiroshima-u.ac.jp}

% \author{
% Ryo~Yamazaki\altaffilmark{1} 
% %% Takayuki~Umeda\altaffilmark{2}, 
% }
% %
% \altaffiltext{1}{Department of Physical Science, Hiroshima University,
% Higashi-Hiroshima 739-8526, Japan; ryo@theo.phys.sci.hiroshima-u.ac.jp}
% %
% %% \altaffiltext{2}{
% %% Solar-Terrestrial Environment Laboratory, Nagoya University, 
% %% Nagoya, 464-8601, Japan; umeda@stelab.nagoya-u.ac.jp}


\def\d{{\rm d}}
\def\p{\partial}
\def\w{\wedge}
\def\o{\otimes}
\def\f{\frac}
\def\tr{{\rm tr}}
\def\Half{\frac{1}{2}}
\def\half{{\scriptstyle \frac{1}{2}}}
\def\T{\tilde}
\def\RA{\rightarrow}
\def\N{\nonumber}
\def\n{\nabla}
\def\bb{\bibitem}
\def\BE{\begin{equation}}
\def\EE{\end{equation}}
\def\BEA{\begin{eqnarray}}
\def\EEA{\end{eqnarray}}
\def\L{\label}
\def\zero{{\scriptscriptstyle 0}}

\begin{abstract}
The two-component emission model to explain the plateau phase
of the X-ray afterglows of gamma-ray bursts (GRBs) is proposed.
One component, which
is responsible for the plateau and subsequent normal decay phase
 of the X-ray afterglow, is the prior emission via outflow ejected from the
 central engine before the main burst.
The other is the main outflow, which causes the 
prompt GRB emission and the initial steep decay phase of the X-ray
 afterglow.
In this model, the transition from the plateau to the 
subsequent normal decay phase
is an artifact of the choice of the zero of time.
For events with distinct plateau phase,
the central engine is active
$10^{3}$--$10^4$~sec before the launch of the main outflow.
According to this model, a prior emission in the X-ray and/or
optical bands 
$10^{3}$--$10^4$~sec before the prompt GRB emission
is possibly seen, which will be
 tested by near-future instruments such as
Monitor of All-sky X-ray Image (MAXI),
WIDe-field telescope for GRB Early Timing (WIDGET), and so on.

\end{abstract}



\section{Introduction}
\label{sec:intro}

The X-Ray Telescope (XRT) onboard {\it Swift}  has
revealed complex temporal behavior of the X-ray afterglows
of gamma-ray bursts (GRBs) in the first few hours
\citep{burrows05,tagl05,nousek06,obrien06,will07,binbin07,liang07,liang08}.
This time window had been largely unexplored before the
{\it Swift} era, and studies of early afterglows have revealed
many questions concerning GRBs, such as the emission mechanism,
nature of the central engine, and burst environment
\citep[e.g.,][]{zhang06,zhang07}.

Early X-ray afterglows have three phases\footnote{
For simplicity, the X-ray flares are not considered in this Letter.
}, which was not predicted
by the standard model from the pre-{\it Swift} era
\citep{nousek06,zhang07}.
{\it Phase~I: initial steep decay phase}.
Initially, the X-ray afterglow
 decays very steeply; the most popular interpretation
is that this is the tail emission of the prompt GRB
\citep{kumar00ng,zhang06,yama06,binbin08}, 
although other possibilities have been proposed 
\citep[e.g.,][]{binbin07}.  
{\it Phase~II: plateau phase}.
At several hundreds of
seconds after the burst trigger, this phase begins until
$\sim10^3$--$10^4$ sec, whose origin is quite uncertain.
This is the main topic of this Letter.
{\it Phase~III: normal decay phase}.
After the plateau phase ends, the X-rays subsequently decay
with the decay index usually steeper than unity, as expected in
the pre-{\it Swift} era. This decay behavior is well explained 
by the classical external shock model \citep{sari98}, in which 
neither the delayed energy injection nor the time dependency of shock 
microphysics is considered.  
%% Hence this phase is sometimes called {\it the normal decay phase}.  
%

Phase~II is the
 most enigmatic in early X-ray afterglows.
So far, various kinds of models have been proposed such as
the energy injection model \citep{nousek06,zhang06,granot06},
the inhomogeneous or two-component jet model 
\citep{toma06,eichler06,granot06m},
the time-dependent microphysics model \citep{ioka06,granot06m,fan06},
the reverse shock-dominated afterglow model \citep{genet07},
the prior activity model \citep{ioka06},
the internal engine model \citep{ghise07},
the cannonball model \citep{dado2006},
and the dust scattering model \citep{shao08}.
In this Letter, another model of phase~II is proposed.

\section{Two-component Emission Model}
\label{sec:model}

\begin{figure}[t]
\includegraphics[width=0.9\textwidth]{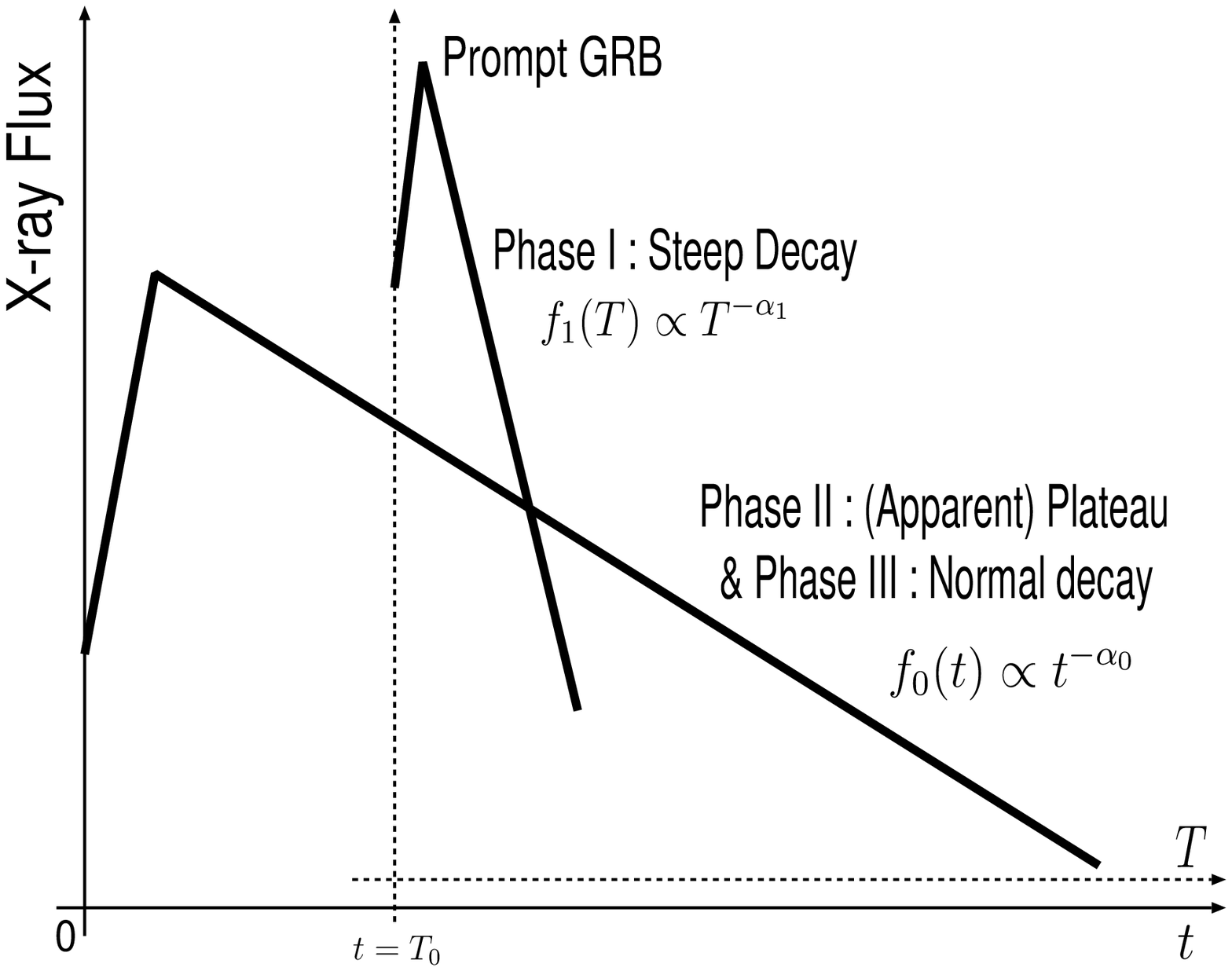}
\caption{
Schematic diagram of the model presented in this Letter.
In terms of the coordinate $t$, the prior emission component
$f_0(t)$ takes a single power-law form throughout the burst,
but in terms of $T(=t-T_0)$, the function $f_0(T)$ has
an artificial plateau phase (see the text for details).
}
\label{fig:model}
\end{figure}

As described in the following,
we consider a two-component emission model in which a
 prior and the main outflows emit X-rays independently
 (see fig.~\ref{fig:model}).
One component is the prior emission via outflow ejected from the
 central engine before the main burst.
This is responsible for phases II and III of the X-ray afterglow.
This emission component arrives at the observer
 before the main burst  triggering prompt
 GRB detectors (i.e., the onset of GRB) such as BAT onboard {\it Swift}.
It decays with time simply in  a single power-law form
\begin{equation}
f_0(t)=A_0 t^{-\alpha_0} ~~,
\label{eq:comp0}
\end{equation}
where $A_0$ is a constant 
and the time coordinate, $t$, is measured in the rest frame
of the observer.
%%  (neglecting cosmic expansion effect for simplicity).
The epoch $t=0$ is taken around the time of arrival at the observer
of the (unseen)
information of the launch of the outflow at the central engine.
This kind of a choice of the time zero is  seen in many references
\citep[e.g.,][]{pana00,kobayashi07}.
The origin of the prior emission, $f_0$, is not discussed in detail
here.
It can be either internal engine activity or external shock emission
of the outflow.

We set an observer time, $T$, where $T=0$ corresponds to the onset of
the prompt GRB. 
The interval between the time $t=0$ and $T=0$ is assumed 
to be $T_0$~seconds, that is,
$t=T+T_0$.
%\begin{equation}
%t=T+T_0 ~~.
%\label{eq:time}
%\end{equation}
%
Then, one obtains
\begin{equation}
f_0(T)=A_0 (T+T_0)^{-\alpha_0} ~~,
\label{eq:comp0_mod}
\end{equation}
which becomes constant if
$T\ll T_0$, while $f_0\propto T^{-\alpha_0}$ when $T\gg T_0$.
In order to explain phases~II and III of the X-ray afterglow,
$T_0$ should be $10^3$--$10^4$~sec, and
$\alpha_0$ should be the temporal decay index of
 phase~III.
It is noted that the onset time of the prior emission is unknown.
The first detectable X-rays from the prior component
arrive at the observer in the time range
$0\lesssim t\lesssim T_0$ ($-T_0\lesssim T\lesssim0$).
This fact will be further discussed in \S~\ref{subsec:precursor}.
Another remark is that
 from Eq.~(\ref{eq:comp0_mod}) alone,
the introduction of $T_0$  shifts the origin of time.
It has been known that, for phase~I, the choice of time zero
affects the decay slope \citep{zhang06,yama06}.
The same argument for phase~II is used
 for the first time in this Letter.

The other component is the main outflow, which causes the 
prompt GRB emission and the subsequent phase~I of the X-ray afterglow.
With time coordinate $T$, 
phase~I of the X-ray afterglow is well approximated
by a single power-law model given by
\begin{equation}
f_1(T)=A_1T^{-\alpha_1}~~,
\label{eq:comp1}
\end{equation}
where $A_1$ is a constant and $\alpha_1\approx3-6$.

The whole light curve of the X-ray afterglow from 
phase~I to III is described by the sum of the two
components introduced above, that is,
\begin{eqnarray}
f(T) &=& f_0(T) + f_1(T) \nonumber\\
     &=& A_0(T+T_0)^{-\alpha_0} + A_1T^{-\alpha_1} ~~.
\label{eq:compall}
\end{eqnarray}
This  is our  formula for
observed X-ray afterglows,
where $f_0(T)$ describes
phases~II and III, while $f_1(T)$ fits phase~I.

\begin{figure}[t]
\includegraphics[width=0.9\textwidth]{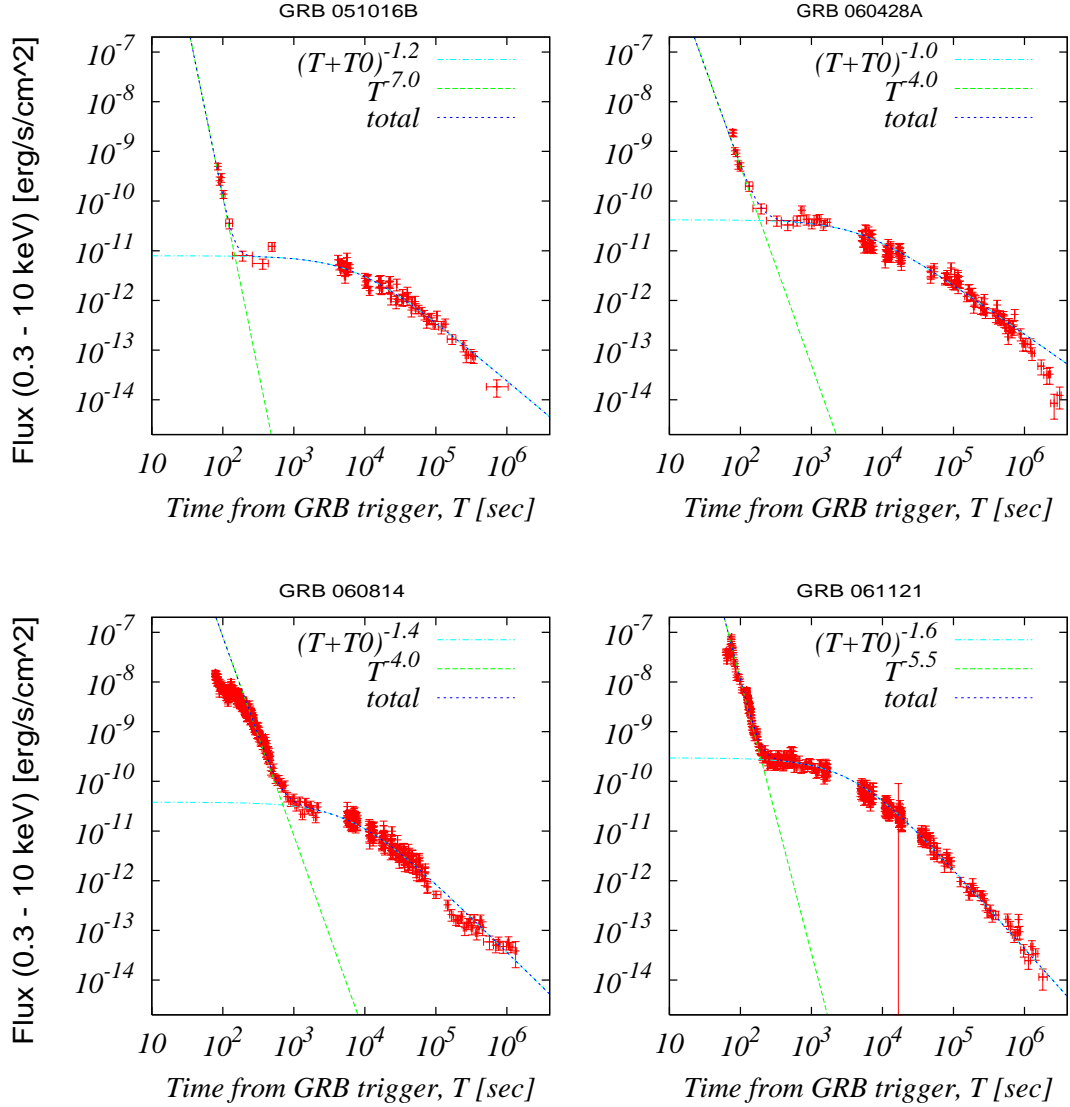}
\caption{
Comparison of the observed light curves of X-ray afterglows with
Eq.~(\ref{eq:compall}), where
all phases I, II, and III are well described.
The adopted parameters are
$(\alpha_0,~\alpha_1,~T_0)=(1.2,~7.0,~8000~{\rm sec})$,
$(1.0,~4.0,~5000~{\rm sec})$,
$(1.4,~4.0,~7000~{\rm sec})$, and
$(1.6,~5.5,~4000~{\rm sec})$, for
GRB~051016B, 060428A, 060814, and 061121, respectively.
Data of X-ray afterglows are taken from the
 {\it Swift} online repository \citep{evans07}.
%% (http://www.swift.ac.uk/xrt\_curves/)
}
\label{fig:fit}
\end{figure}

We find that by choosing appropriate values of the parameters,
$\alpha_0$, $\alpha_1$, $T_0$, $A_0$, and $A_1$, 
the observed light curves of  X-ray afterglows
are well described with Eq.~(\ref{eq:compall}).
Figure~\ref{fig:fit} shows some examples of the fit.

It is not surprising that our formula, Eq.~(\ref{eq:compall}),
well explains the observational results of X-ray afterglows.
The functional form of Eq.~(\ref{eq:comp0_mod}) is a good approximation of
that introduced by \citet{will07}:
\begin{equation}
f_a(T)=
\left\{
\begin{array}
{l@{~~} c@{}}
F_a \exp[\alpha_a(1-T/T_a)]\exp(-t_a/T)~, & T<T_a~, \\
F_a(T/T_a)^{-\alpha_a}\exp(-t_a/T)~, & T_a\leq T~.
\end{array} \right. 
\label{eq:will}
\end{equation}
They have shown that the observed light curves of phases~II and III
are well fitted with Eq.~(\ref{eq:will}).
The function $f_a(T)$ becomes constant with $T$ if $t_a\ll T\ll T_a$,
while $f_a\propto T^{-\alpha_a}$ if $T_a\ll T$.
This behavior is quite similar to that of $f_0(T)$.
Indeed, one can find that over a wide parameter range,
$f_a(T)\approx f_0(T)$ for $t_a\ll T$ if we take
$t_a\ll T_a\approx T_0$ and $\alpha_a\approx\alpha_0$.

\section{Discussion}
\label{sec:discussion}

\subsection{Overall shape of the X-ray afterglow}

In this Letter, we have seen
that for most events,  phases~II and III of the
observed light curves of the X-ray afterglow can be well
fitted with a very simple formula, Eq.~(\ref{eq:comp0_mod}).
In particular, the observational facts are that
for most events, the transition from phase~II to III
is slow and fairly smooth, and that
the X-ray spectrum remains unchanged across the
transition \citep{nousek06};
this can be naturally explained by our model
because the transition from phase~II to III is an artifact of
the choice of the time zero.
A typical exception is GRB~070110;
at the end of the plateau phase~II, the light curve shows an
abrupt drop
%%, which rules out an external shock origin
 \citep{troja07}.
Such events may need other explanations \citep[e.g.,][]{kumar08}.

After  phase~III, subsequent fourth phase is occasionally observed:
the so called {\it post jet break phase} 
\citep[{\it Phase~IV};][]{zhang07}.
Its typical decay index is $\sim-2$, satisfying the predictions
of the jetted afterglow model \citep{sari99}.
This phase~IV can also be clearly seen  in our sample
(e.g., GRB~060428A; upper right of Fig.~\ref{fig:fit}).
If the prior emission is from an external shock of the prior outflow,
the jetted afterglow model is applied to the present case.
As will be discussed in \S~\ref{subsec:MainAG},
the origin of the optical afterglow is different from the X-ray one,
which explains the fact that the epoch of the jet break in the 
X-ray band is not generally the same as the optical one 
\citep{sato07,liang08}.

\subsection{Observed $f_0(T_0)$ -- $T_0$ Correlation}

\citet{sato08} investigated the characteristics of the transition from
phase~II to III for 11 events with known redshifts.
They derived the transition time, $T^0_{\rm brk}$, and
the isotropic luminosity at that time $L_{\rm X, end}$,
and found that 
$L_{\rm X, end}$ is well correlated with
$T^0_{\rm brk}$ as $L_{\rm X, end}\propto (T^0_{\rm brk})^{-1.4}$.
They adopt a broken power-law form to fit the
light curve in  phases~II and III which is different from that
considered in this Letter. However,
we expect that their $T^0_{\rm brk}$ and $L_{\rm X, end}$
roughly correspond to $T_0$ and $f_0(T_0)$, respectively, 
that are considered in \S~\ref{sec:model}.
Indeed, one finds 
$f_0(T_0)\propto T_0{}^{-\alpha_0}$ from Eq.~(\ref{eq:comp0_mod}).
Hence, if $\alpha_0\approx1.4$ which is a typical number for
the decay index of  phase~III, we can reproduce the observed result
of \citet{sato08}.

A similar analysis has been done but with
Eq.~(\ref{eq:will}) as the fitting formula \citep{dainotti08}.
For 32 events with measured redshifts,
they found a correlation between $T_a$ and
the X-ray luminosity at the time $T_a$, $L_X(T_a)$,
 as $L_X(T_a)\propto T_a^{-\beta}$ with $\beta=0.6$--0.74.
Since $T_a\approx T_0$ and $L_X(T_a)\approx f_0(T_0)$,
their correlation indicates $f_0(T_0)\propto T_0{}^{-\beta}$.
The index $\beta$ is smaller than the typical value of 
$\alpha_0$ ($\approx1.0$--1.5).
However, the claimed correlation has large 
scatter \citep[see Figures~1 and 2 of][]{dainotti08},
which may be explained by the scatter of $\alpha_0$ in our model.

There is a difference between the results of \citet{sato08}
and \citet{dainotti08}.
At present, the number of events with known redshifts may be small,
so this discrepancy may be resolved if the number of events
increases.

\subsection{Link between $T_0$ and the Prompt Emission Properties}

\citet{nava07} studied the properties of prompt emission 
and X-ray afterglows of 23 GRBs with known redshifts.
They adopted Eq.~(\ref{eq:will}) in fitting phases~II and
III of the X-ray afterglow, and found that for events with
measured spectral peak energy $E_{\rm p}$, the time
$T_a$ weakly correlates with the isotropic equivalent energy 
$E_{\gamma,{\rm iso}}$ of the prompt GRB emission.
One can find from Fig.~6 of \citet{nava07},
that $T_a$ seems to be roughly proportional to $E_{\gamma,{\rm iso}}$.
At present, this correlation is not firmly established
because as noted by \citet{nava07}, there are no correlation
between $T_a$ and the isotropic equivalent energy of 
prompt GRBs in the 15-150~keV band 
for a larger sample of GRBs with known redshift but unknown $E_{\rm p}$
(hence without $k$-correction).

The quantity $E_{\gamma,{\rm iso}}$ correlates with $E_{\rm p}$
\citep{amati02}.
Hence, if the  $T_a$--$E_{\gamma,{\rm iso}}$ correlation exists,
 the bright GRBs with large $E_{\gamma,{\rm iso}}$
and $E_{\rm p}$ have large $T_a$, which is responsible for 
the distinct plateau phase. On the other hand,
the X-ray flashes or X-ray-rich GRBs 
\citep[e.g.,][]{heise01,barraud03,sakamoto08}
have small $T_a$ ($\approx T_0$),
and have X-ray afterglow without  phase~II.
This tendency could have been seen in \citet{sakamoto08}.

The above arguments may lead a link between long GRBs
and X-ray flashes/X-ray-rich GRBs.
Suppose that the outflow ejection is not continuous but
intermittent, i.e.,
the central engine ejects two distinct outflows with a
time interval of $\sim T_0$.
Just after the launch of the prior outflow, 
the central engine does not have enough energy  for
another outflow, so that
it  needs to store  an additional one.
During the time interval $\sim T_0$,
matter surrounding the central engine
is accreted, increasing the gravitational binding energy.
This energy is released as the main outflow causing the prompt GRB.
It is expected that the larger is $T_0$,
the larger is the stored gravitational energy, resulting in a
 brighter burst with large $E_{\gamma,{\rm iso}}$.
On the other hand, if $T_0$ is small the energy of the main
outflow becomes small; this is responsible for the X-ray flash
or X-ray-rich GRBs.
Further details will be discussed in the near future.

\subsection{External Shock Emission from the Main Outflow}
\label{subsec:MainAG}

The main outflow that is responsible for the prompt GRB 
might cause external shock X-ray emission, $f_{\rm X, ext}(T)$.
In the present two-component emission model, however,
$f_{\rm X, ext}(T)$ must be dimmer than the
prior X-ray emission $f_0(T)$ throughout  phases~II and III.
Let us consider the simplest model of  external shock emission of
 the main outflow.
The relativistically expanding shell with energy $E_K$ interacts with 
the surrounding medium with uniform density $n_0$\footnote{
Since the prior outflow may modify the circumburst medium
density profile, the external shock emission from the main 
outflow deviates from the case of a uniform density profile. 
Nevertheless, we adopt the uniform density model here
for the simplicity.}, 
and emits synchrotron
radiation with microphysics parameters at the shock,
$p$, $\varepsilon_e$, and $\varepsilon_B$ \citep{sari98}.
Then, in the case of slow cooling and $\nu_c<\nu_X$, 
the X-ray light curves are analytically
calculated as 
$f_{\rm X, ext}(T) \propto
E_K^{(p+2)/4} \varepsilon_e^{p-1} 
\varepsilon_B^{(p-2)/4} T^{(2-3p)/4} \nu_{X}^{-p/2}$,
%\begin{equation}
%f_{\rm X, ext}(T) \propto
%E_K^{(p+2)/4} \varepsilon_e^{p-1} 
%\varepsilon_B^{(p-2)/4} T^{(2-3p)/4} \nu_{X}^{-p/2}~~,
%\end{equation}
which is independent of $n_0$ \citep[e.g.,][]{pana00}.
If $p\approx2$, then $f_{\rm X, ext}(T)$ hardly depends on 
 $\varepsilon_B$.
Before the {\it Swift} era, typical values had been
$E_K\sim10^{52}$--$10^{53}$~erg, 
%% $n\sim1$--10~cm$^{-3}$,
$\varepsilon_e\sim10^{-1}$, and $\varepsilon_B\sim10^{-2}$
so that the external shock emission reproduced the observed late-time
X-ray afterglow.
In the present case, $f_{\rm X, ext}(T)$ must be dim, which implies
small $E_K$ and/or $\varepsilon_e$.
A similar discussion was made by \citet{ghise07,ghise08}.
In some models, such as the energy injection model
and the inhomogeneous jet model, prompt GRB emission
needs high radiation efficiency, which is defined by
$\varepsilon_\gamma=E_{\gamma,{\rm iso}}/(E_{\gamma,{\rm iso}}+E_K)$,
because $E_K$ is small at the epoch of the prompt GRB emission
\citep{fan06,granot06m,ioka06,zhang07effi}.
As discussed here, $f_{\rm X, ext}(T)$ can be dim
 if $\varepsilon_e$ is small while $E_K$ remains 
large, $\sim10^{52}$--$10^{53}$~erg.
Hence, the present model could avoid a serious
efficiency problem.
On the other hand, if $E_K$ of the outflow is small,
the efficiency $\varepsilon_\gamma$ should be high.
Then the mechanism of  prompt GRB emission
is unlike a classical internal shock model
\citep[e.g.,][]{thompson07,ioka07}.

The observed optical afterglow  comes mainly from the prior outflow.
For some events, the rising part of the early optical afterglow proceeds
until $T\sim10^2$~sec \citep{moli07}, which is difficult to  
explain  with  prior emission.
In the case of  prior emission,
the time zero would be shifted $T_0\sim10^3$--$10^4$~sec before the 
burst trigger, making the light curve extraordinarily spiky.
Furthermore, in most cases, the transition from phase~II to III is 
chromatic, i.e., 
the optical light curves do not show
any break at that epoch \citep{pana06}, although there exist
a few exceptions \citep{liang07,grupe07,mangano07}.
Hence, at least in the early epoch,
the observed optical afterglow arises from the main outflow 
component or
others,  most likely an external shock emission.
Indeed, there have been some observational facts that indicate 
different origins of X-ray and optical afterglows 
\citep[e.g.,][]{oates07,sato07,urata07}\footnote{
One may expect that the optical emission also arises
from the prior outflow.
However, it may be outshone by the
main outflow component at least in the early epoch. 
This condition will constrain the
mechanisms of the prior X-ray  emission as well as the optical one.
}.

\subsection{Predicted Precursor Emission?}
\label{subsec:precursor}

A possible prediction of the present model is a bright X-ray precursor
before the prompt GRB emission.
Let us assume that the prior X-ray emission starts at
$t\sim10^2$~sec, although its onset time is fairly uncertain
(see \S~\ref{sec:model}).  Then, from Eq.~(\ref{eq:comp0}),
the X-ray flux in the 2--10~keV band is estimated as
$f_0(t)\sim 4\times10^{-9}(t/10^2~{\rm s})^{-1.2}$erg~s$^{-1}$cm$^{-2}$,
%\begin{equation}
%f_0(t)\sim 4\times10^{-9}\left(\f{t}{10^2~{\rm s}}\right)^{-1.2}
%{\rm erg}~{\rm s}^{-1}{\rm cm}^{-2}~~,
%\end{equation}
where $\alpha_0\approx1.2$ is taken as a typical value and
the flux normalization constant $A_0$ is determined so that
$f_0\sim1\times10^{-12}$~erg~s$^{-1}$cm$^{-2}$ at
$t\sim10^5$~sec \citep{gendre08}.
Such  emission will be detected by 
Monitor of All-sky X-ray Image (MAXI)\footnote{
http://www-maxi.tksc.jaxa.jp/indexe.html
}.
The expected event rate should be a few events per year
\citep{motoko08}.

However, if the emission starts with $t\sim10^2$s, the predicted
flux might be large enough to be detected by current instruments
like BAT onboard {\it Swift}.
In order to avoid this problem, the starting time of the emission 
should be comparable to or later
 than $t\sim10^3$~sec so that the peak flux is smaller
than the detection limit of the prompt GRB emission monitors.
One possibility is  off-axis jet emission in the context of 
the external shock model \citep{granot03,granot05}.
Due to the relativistic beaming effect, the observed X-ray emission
is dim as long as the bulk Lorentz factor of the emitting outflow
is larger than the inverse of the angle between the emitting matter
and the observer's line of sight.
This effect shifts the peak time of the X-ray emission toward the
later epoch.
%
%Another possibility to delay the onset of the prior X-ray emission
%is the opacity effect, which can be expected in the context of 
%the internal dissipation (ref).
%Initially the emission occurrs at a small radius in which the pair
%formation prevents the photons from escaping the emitting region.
%
For an appropriate choice of parameters, we may adjust the
starting time (the peak time) of the X-ray emission.

A signal of the onset of the prior emission might also be seen in the
optical band.
If the prior emission is extremal shock origin,
the reverse shock emission might cause a bright optical flash
\citep{sari99opt}.
So far, for some events, WIDGET\footnote{
http://cosmic.riken.go.jp/grb/widget/}
has given observational upper
limit on the prior optical emission, $V>10$~mag, about 750~seconds
before the prompt emission \citep[e.g.,][]{abe06}.
Further observations will constrain the model parameters.
In summary,  in order to test the model presented in this Letter,
the search for a signal in the data of sky monitors in
the optical and X-ray bands is crucial.

\acknowledgments

R.Y. would like to thank 
Takashi~Nakamura, Kunihito~Ioka, Takanori~Sakamoto,
Atsumasa~Yoshida, Motoko~Suzuki, 
Takeshi~Uehara, and the anonymous referee
 for useful comments and discussions.
%% and the anonymous referee for useful comments.
%
This work was supported in part 
by grant-in-aid from the 
Ministry of Education, Culture, Sports, Science,
and Technology (MEXT) of Japan,
No.~18740153, No.~19047004.
%% (R.~Y.).

\clearpage

\end{document}